\documentclass[11pt]{article}
\usepackage{hyperref}
\pdfoutput=1
\begin{document}
\title{Liquid acrobatics}
\author{James C. Bird and Howard A. Stone \\
\\\vspace{6pt} School of Engineering and Applied Sciences, \\ Harvard University, Cambridge, MA 02138, USA}
\maketitle
\begin{abstract}
We experiment with injecting a continuous stream of gas into a shallow liquid, similar to how one might blow into a straw placed at the bottom of a near-empty drink.  By varying the angle of the straw (here a metal needle), we observe a variety of dynamics, which we film using a high-speed camera.  Most noteworthy is an intermediate regime in which cyclical jets erupt from the air-liquid interface and breakup into air-born droplets.  These droplets trace out a parabolic trajectory and bounce on the air-liquid interface before eventually coalescing.  The shape of each jet, as well as the time between jets, is remarkably similar and leads to droplets with nearly identical trajectories.  The following article accompanies the linked fluid dynamics video ( \href{http://ecommons.library.cornell.edu/bitstream/1813/11469/3/Bird_DFD2008_mpeg1.mpg}{low resolution},
 \href{http://ecommons.library.cornell.edu/bitstream/1813/11469/2/Bird_DFD2008_mpeg2.mpg}{high resolution})
 submitted to the Gallery of Fluid Motion in 2008.
\end{abstract}

\section{Introduction}

This paper documents some of the dynamics observed when a gas is injected into a shallow pool of liquid.  These dynamics are visible to the naked eye, but benefit from the high-frame rate and low exposure time offered by a high-speed camera.  The high-speed video discussed in this paper is available in the accompanying \href{http://ecommons.library.cornell.edu/bitstream/1813/11469/3/Bird_DFD2008_mpeg1.mpg}{movie}, submitted to the 2008 American Physical Societies Division of Fluid Dynamic's Gallery of Fluid Motion.

\section{Setup}
The experimental setup consists of a 7 gauge metal needle placed into a 5-millimeter deep pool of 10 cSt silicon oil, such that part of the needle tip contacts the bottom of the container.  Similar regimes have been observed in other liquids, although the dynamics are less crisp.  Helium is pumped through the needle at a modest flowrate of 3 mL/s.  The advantage of helium is that it is well-characterized, inert, and readily available; we suspect that the dynamics shown in the movie are independent of the gas properties.

The flow of helium into the silicon oil distorts the interface between the silicon oil and surrounding air.  The precise dynamics of this deformation depend on the angle of needle, flowrate of the gas, and the depth and material properties of the liquid.  In this study we have held the flowrate and pool depth constant while varying the angle of the needle, $\theta$.

\section{Results and discussion}
At low needle angles ($\theta < 57^\circ$), the air-silicon oil interface is in static equilibrium.  Once the angle approaches $57^\circ$, ripples propagate across the surface.  Due to the size and speed of the ripples, we believe that they are capillary waves, that is waves that are regulated by surface tension.  As the angle of the needle continues to increase, the amplitude of the capillary waves grows such that an oscillating bulge develops.  At higher angles, the bulge becomes both larger and more focused, developing into periodic jets.  Once the jet aspect ratio of the jet becomes sufficiently small ($\theta = 66^\circ$), the tip of the jet pinches off to form a spherical droplet.

The droplets formed from the periodic jets float over the interface, similarly to how drops avoid coalescence over a vibrating fluid \cite{cou05}.  It is worthwhile to note that the droplets often coalesce with each other before coalescing with the interface, implying that it is unlikely that natural surfactants are responsible for the floating behavior.  As the angle of the needle increases, the height of the upward jets increase.  At sufficient jet height ($\theta \approx 72^\circ$), the jets break into two drops, leading to both a high parabolic arch and a lower droplet stream.  Between ($80^\circ < \theta < 90^\circ$), the jet dynamics breakdown so that the droplet streams appear chaotic.  Large bubbles sporadically form at the fluid interface, often with jets emitting from the bubble surface.  Droplets are still emitted from the jets, but the resulting bubble streams are incoherent.

\section{Conclusion}
We have found that injecting gas into shallow liquid can lead to a rich variety of interfacial dynamics.  By sweeping the angle of the gas-injecting needle, we observe a smooth transition between static deformation, juggling and tumbling of droplet streams, and an incoherent mix of bubbles and jets, reminiscent of a daredevil shooting out of a canon.  This combination of dynamics provides an aesthetically pleasing show of liquid acrobatics.


\begin{thebibliography}{1}
\bibitem{cou05}
Y. Couder, S. Protiere, E. Fort and A. Boudaoud, Nature {\bf 437},    (2005).
\end{thebibliography}
\end{document}